\documentclass[journal]{IEEEtran}
\usepackage{hyperref}
\usepackage{amsthm}
\usepackage{stfloats} 
\usepackage{array,cite}
\usepackage{pgfplots}
\usepgfplotslibrary{units}
\usetikzlibrary{patterns}
\usepackage{multirow} 
\usepackage{stmaryrd}
\usepackage{tikz}
\usepackage{booktabs}
\usetikzlibrary{arrows,calc,positioning}
\usepackage{footnote}
\usepackage{amsmath,amssymb}
\usepackage{makecell}
\usepackage{float}
\usepackage[ruled]{algorithm2e}
\usepackage{algorithmic}
\pgfplotsset{compat=1.14}

\definecolor{IvoryBlack}{rgb}{0.1600, 0.1400, 0.1300}
\definecolor{GreenDark}{rgb}{0,0.3922,0}
\definecolor{RedLight}{rgb}{0.8922,0,0}
\definecolor{BlueLight}{rgb}{0, 0.3, 0.6039}
\definecolor{OrangeLight}{rgb}{0.9500, 0.6600, 0.4400}
\definecolor{VioletLight}{rgb}{0.5765, 0.4392, 0.8588}
 
 \newcommand{\fixme}[2]{\ifx&#2&{\color{red}#1}\else{\color{red}FIXME\{}#1{\color{red}\}}\footnote{{\color{red}#2}}\PackageWarning{Fixme}{#1: #2}\fi}

\DeclareMathOperator*{\argmax}{arg\,max}

\begin{document}

\title{Implementation of a High-Throughput Fast-SSC Polar Decoder with Sequence Repetition Node}

\author{\IEEEauthorblockN{Haotian~Zheng, Alexios~Balatsoukas-Stimming, Zizheng~Cao, Ton~Koonen}
\IEEEauthorblockA{\\Department of Electrical Engineering, Eindhoven University of Technology, 5612 AZ Eindhoven, The Netherlands\\
\{h.zheng, a.k.balatsoukas.stimming, z.cao, a.m.j.koonen\}@tue.nl}
\vspace{-0.5em}
}

\maketitle

\begin{abstract}
Even though polar codes were adopted in the latest 5G cellular standard, they still have the fundamental problem of high decoding latency. Aiming at solving this problem, a fast simplified successive cancellation (Fast-SSC) decoder based on the new class of sequence repetition (SR) nodes has been proposed recently in~\cite{sr2020} and has a lower required number of time steps than other existing Fast-SSC decoders in theory. This paper focuses on the hardware implementation of this SR node-based fast-SSC (SRFSC) decoder. 
The implementation results for a polar code with length 1024 and code rate $1/2$ show that our implementation has a throughput of $505$ Mbps on an Altera Stratix IV FPGA, which is $17.9\%$ higher with respect to the previous work.
\end{abstract}

\IEEEpeerreviewmaketitle

\section{Introduction}
\label{sec:intro}

Polar codes are the first provably capacity-achieving channel codes with an explicit construction, low-complexity encoding and decoding algorithms, and easily adaptable coding rate~\cite{arikan2009}. Although the capacity of binary symmetric memoryless channels can be achieved using the low-complexity successive cancellation (SC) decoding algorithm, the sequential nature of SC decoding typically leads to a large decoding latency, which constrains its application in high-throughput and low-latency communication scenarios such as 5G and optical wireless communications~\cite{koonen2020}. A simplified successive cancellation (SSC) decoder was proposed in~\cite{alamdar2011simplified}, where fast decoding methods are described for subcodes of the polar code (called \emph{constituent codes}) that consist either of only information bits or of only frozen bits. Following this idea, other constituent codes with special information bit patterns and their corresponding fast decoders were identified in \cite{sarkis2014fast,hanif2017fast,condo2018generalized,gamage2019low}. The family of these decoding algorithms is often referred to as \emph{fast-SSC} decoding. To increase the number of fast decoding constituent codes, \cite{638mbps,giard2018fast} altered
the polar code construction to further improve latency at the cost of a small error-correcting performance degradation. Methods to optimize the memory footprint of fast-SSC decoders were described in~\cite{Furkan2017}.

The work of \cite{sr2020} proposed a new class of \emph{sequence repetition } (SR) constituent codes, which is a generalization of most existing constituent codes. It was also shown that the decoding of SR constituent codes can be highly parallelized to achieve further latency reduction compared to the state of the art without tangibly affecting the error-correcting performance. However, the work of \cite{sr2020} only focused on the algorithmic aspects of SR constituent codes and no hardware implementation has been reported in the literature.

\subsubsection*{Contribution} In this work, we describe a hardware architecture for a fast-SSC decoder that exploits the SR constituent codes described in \cite{sr2020} and we provide FPGA implementation results. Even though our proposed implementation is not yet highly optimized, it still achieves a $17.9$\% higher decoding throughput than the state of the art.

\section{Background}
\label{sec:pre}

\subsection{Polar Codes}
\label{sec:PC}
A polar code with code length $N=2^n$ and information length $K$ is denoted by $\mathcal{P}\left(N,K\right)$ and has rate $R = K/N$.
The input bit sequence $\boldsymbol{u}$ consists of $K$ information bits whose positions form set $\mathbb{A}$ and $N-K$ frozen bits whose positions form $\mathbb{A}^c$. The values of the frozen bits are usually set to $0$. The encoded bit sequence can be calculated as $\boldsymbol{x}=\boldsymbol{u}\mathbf{G}_N$, where $\mathbf{G}_N=\mathbf{R}_N\mathbf{F}_2^{\otimes n}$ is the generator matrix of the polar code, $\mathbf{R}_N$ is a bit-reversal permutation matrix and $\mathbf{F}_2=\left[\begin{smallmatrix}1&0\\1&1\end{smallmatrix}\right]$.

\subsection{SC and Fast-SSC Decoding}
\label{sec:FSC}

\subsubsection{Algorithm}

\begin{figure}
\centering
\scalebox{0.75}{\includegraphics{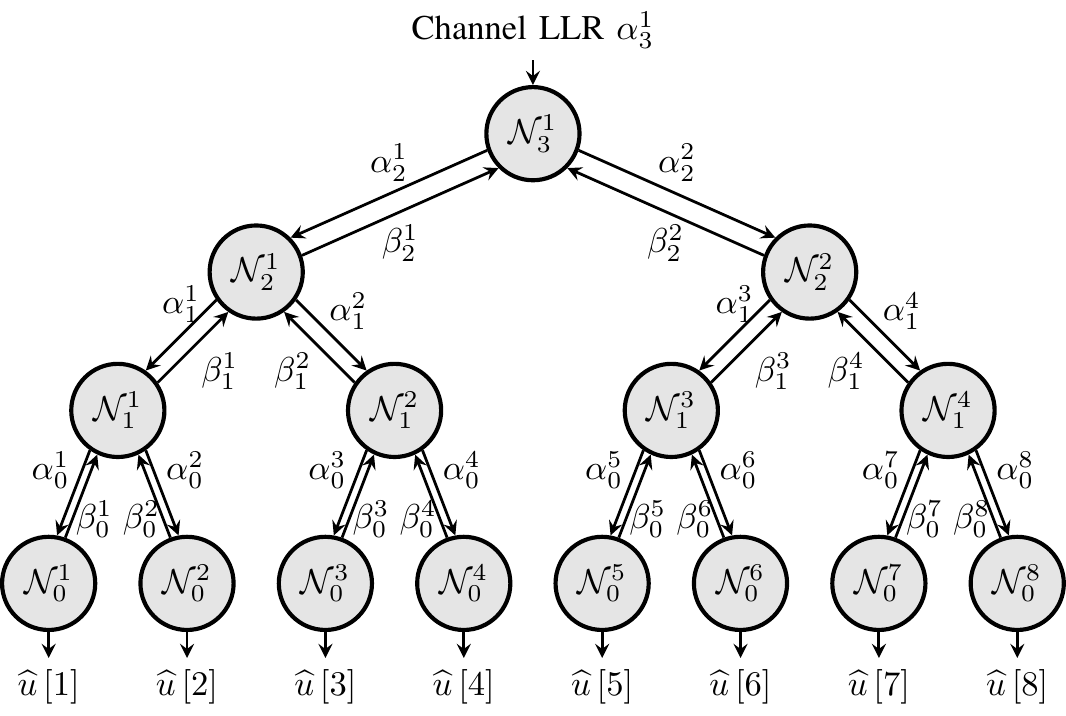}}
\caption{SC decoding tree representation of a polar code with $N=8$.}
\label{fig:tree}
\end{figure}

SC decoding of polar codes can be represented as the traversal of a binary tree as in Fig.~\ref{fig:tree}. The $i$-th node at level $j$ ($1\leq i\leq 2^{n-j}$) of the SC decoding tree corresponds to a constituent code with bit index from $2^j\cdot\left(i-1\right)+1$ to $2^j\cdot i$, and is denoted as $\mathcal{N}_j^i$. The left and the right child nodes of $\mathcal{N}_j^i$ are $\mathcal{N}_{j-1}^{2i-1}$ and $\mathcal{N}_{j-1}^{2i}$, respectively. For $\mathcal{N}_j^i$, the symbol $\alpha_j^i\left[k\right]$, $1\leq k\leq2^j$, denotes the $k$-th input logarithmic likelihood ratio (LLR) value, and $\beta_j^i\left[k\right]$, $1\leq k\leq2^j$, denotes the $k$-th output binary hard-valued message. The SC decoding follows a depth-first principle, with priority
to the left branch. When LLR messages pass to the left and right child nodes, $f$ and $g$ functions over the LLR domain are executed, respectively, which are given by
\begin{equation}
\begin{aligned} \label{eq:f_function}
\alpha_{j-1}^{2i-1}\left[k\right]\approx &\text{sign}\left(\alpha_j^{i}\left[{2k-1}\right]\right)\text{sign}\left(\alpha_j^{i}\left[{2k}\right]\right)\\
&\cdot\min\left(\alpha_j^{i}\left[{2k-1}\right],\alpha_j^{i}\left[{2k}\right]\right),
\end{aligned}
\end{equation}
\begin{equation}
\label{eq:g_function}
\alpha_{j-1}^{2i}\left[k\right]=\left(-1\right)^{\beta_{j-1}^{2i-1}\left[k\right]}\alpha_j^{i}\left[{2k-1}\right]+\alpha_j^{i}\left[{2k}\right].
\end{equation}
When the LLR value of the $k$-th bit at level zero $\alpha_0^k,\;1\leq k\leq N$, is calculated, the estimation of $u\left[k\right]$, denoted as ${\hat u}\left[k\right]$, is
\begin{equation}\label{eq:decision}
\centering      
\hat{u}\left[k\right]=\hat\beta_0^k=\begin{cases} 0, &\mbox{if } k\in \mathbb{A}^c \text{,} \\ 
\frac{1-\mathrm{sign}(\alpha_0^k)}{2}, & \mbox{otherwise.} \end{cases}
\end{equation}
The hard messages are propagated back to the parent node as
\begin{equation}
\hat\beta_j^i\left[k\right]=
\begin{cases} \hat\beta_{j-1}^{2i-1}\left[\frac{k+1}2\right]\oplus\hat\beta_{j-1}^{2i}\left[\frac{k+1}2\right], &\mbox{if} \mod(k,2)=1 \text{,} \\ 
\hat\beta_{j-1}^{2i}\left[\frac{k}2\right], & \mbox{if} \mod(k,2)=0. \end{cases}
\label{eq:hardpropagation}
\end{equation}

The estimation of each bit depends on the estimation of all previous bits in the SC decoding algorithm, which leads to a large latency. It was pointed out in \cite{sarkis2013increasing} that for a node $\mathcal{N}_j^i$, the maximum-likelihood (ML) estimate of the vector $\beta_j^i\left[1:2^j\right]$ can be calculated in parallel by evaluating
\begin{equation} \label{eq:estimate}
\hat\beta_j^i\left[1:2^j\right]= \underset{\beta_j^i\left[1:2^j\right]\in\mathbb{C}_j^i}{\argmax}\sum_{k=1}^{2^j}\left(-1\right)^{\beta_j^i\left[k\right]}\alpha_j^i\left[k\right],
\end{equation}
where $\mathbb{C}_j^i$ is the set of all the codewords associated with node $\mathcal{N}_j^i$. The complexity of evaluating~\eqref{eq:estimate} is generally very high. However, the main idea behind fast-SSC decoding is that for some nodes with special frozen and non-frozen bit patterns the evaluation of~\eqref{eq:estimate} can be simplified significantly. Some prominent examples of such nodes include the Rate-0 node ($2^j$ frozen bits), the Rate-1 node ($2^j$ information bits), the repetition (REP) node (one information bit and $2^j-1$ frozen bits), and the single parity-check (SPC) node ($2^j-1$ information bits and one frozen bit). 

The key advantage of using specific parallel decoders for the aforementioned special nodes is that, since the SC decoding tree is not traversed when one of these nodes is encountered, a significant latency reduction can be achieved. For example, if $\mathcal{N}_j^i$ is a Rate-1 node, hard decision decoding can be used to immediately obtain the decoding result as
\begin{equation}
\hat\beta_j^i\left[k\right]=h\left(\alpha_j^i\left[k\right]\right)=
\begin{cases} 0, &\mbox{if } \alpha_j^i\left[k\right]\geq0 \text{,} \\ 
1, & \mbox{otherwise.} \end{cases}
\label{eq:rate1}
\end{equation}
If $\mathcal{N}_j^i$ is a REP node, all its bits are either equal to one or equal to zero. According to \eqref{eq:estimate}, estimation can be obtained by extracting the sign bit of the sum of its LLR values. If $\mathcal{N}_j^i$ is an SPC node, a hard decision based on~\eqref{eq:rate1} is first performed, which is followed by the calculation of the parity of the output using modulo-2 addition. The hard decision value with the index of the least reliable bit will be flipped if the parity check constraint is not met. 

\subsubsection{Fast-SSC Decoder Hardware Architectures}

A typical fast-SSC decoder contains three main modules \cite{sarkis2014fast}: a memory, an arithmetic logical unit (ALU), and a controller. The memory consists of five separate sub-modules. The channel LLR, internal LLR $\alpha$, and estimation $\beta$ sub-modules feed the ALU. The instruction sub-module stores the operations to be executed and is routed into the controller. Finally, the codeword sub-module stores and outputs the final codeword. 
The ALU implements the $f$ function given in \eqref{eq:f_function}, the $g$ function given in \eqref{eq:g_function}, the combining operation given in~\eqref{eq:hardpropagation}, as well as the update rules for various special nodes like the rate-$1$ node given in \eqref{eq:rate1}. 
Finally, the controller tracks which node in the decoding tree is currently being decoded by using a list of instructions that is pre-compiled based on $\mathbb{A}$ and $\mathbb{A}^c$. 

\section{Fast-SSC Decoding with Sequence Repetition Nodes}
\label{sec:sr}
\subsection{Sequence Repetition Node}
\label{sec:sr node}

\begin{figure}
\centering
\scalebox{0.8}{\includegraphics{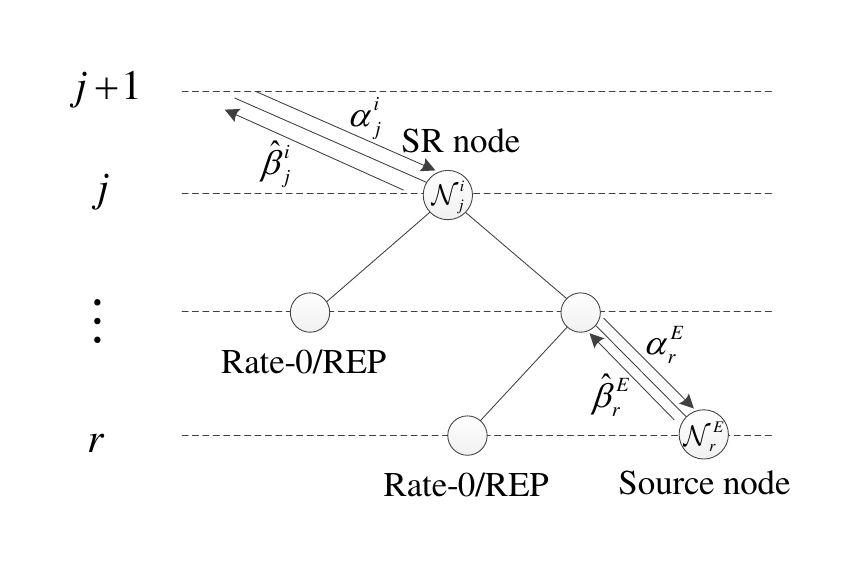}}
\setlength{\abovecaptionskip}{-15pt}
\caption{General structure of a sequence repetition node.}
\label{fig:SR}
\end{figure}

Let $\mathcal{N}_j^i$ be a node at level $j$ of the binary tree representation of SC decoding as shown in Fig.~\ref{fig:tree}. An SR node is any node at stage $j$ for which all its descendants are either Rate-0 or REP nodes, except the rightmost one at a certain stage $r$, $0\leq r\leq j$, that is a generic node of rate $C$. The general structure of an SR node is depicted in Fig.~\ref{fig:SR}.
The rightmost node $\mathcal{N}_r^{i\times 2^{j-r}}$ at stage $r$ is denoted as the source node of the SR node $\mathcal{N}_j^i$. Let $E = i\times 2^{j-r}$ so the source node can be denoted as $\mathcal{N}_r^E$.

An SR node can be represented by three parameters as $\text{SR}(\boldsymbol{v},\text{SNT},r)$, where $r$ is the level of the SC decoding tree in which $\mathcal{N}_r^E$ is located. SNT is the source node type, and as shown in \cite{sr2020}, $\text{SNT}\in\{\text{Rate-0}, \text{Rate-1}, \text{EG-PC}, \text{Rate-C}\}$. The EG-PC node is a node at level $j$ having all its descendants as Rate-1 nodes except the leftmost one at a certain level $r<j$, that is a Rate-0 or REP node. $\text{Rate-C}$ is a generic node of rate $C$. When $\text{SNT}\in\{\text{Rate-0}, \text{Rate-1}, \text{EG-PC}\}$, the source node is a special node whose bits are all non-frozen except the leftmost $b$ bits, where 
\begin{align}
    b = \left\{
        \begin{matrix}
            0, & \text{if\;SNT=Rate-1}, \\
            1, & \text{if\;SNT=Rate-0}, \\
            2^h \text{ or } 2^h-1, & \text{if\;SNT=EG-PC},
        \end{matrix}
    \right. \label{eq:a}
\end{align}
and where $h<r-1$ is the level of the leftmost Rate-0/REP node of the EG-PC node. 
Note that the source node has a minimum length of 2 as all the possible frozen bit patterns with length 2 fall into the above category. The vector $\boldsymbol{v} = \left(v\left[{j}\right],v\left[{j-1}\right],\ldots,v\left[{r+1}\right]\right)$ has length $\left(j-r\right)$ such that for the left child node of the parent node of $\mathcal{N}_r^E$ at level $k$, $r<k\leq j$, $v\left[{k}\right]$ is calculated as
\begin{equation}
v\left[k\right] = \begin{cases}
0, & \text{if the left child node is a Rate-0 node,}\\
1, & \text{if the left child node is a REP node.}
\end{cases}
\end{equation}
Note that when $r=j$, $\mathcal{N}_j^i$ is a source node and thus $\boldsymbol{v}$ is an empty vector denoted as $\boldsymbol{v} = \emptyset$.

\subsection{Repetition Sequence}

In this subsection, we define \emph{repetition sequences}, which can be used to calculate the output bit estimates of an SR node based on the estimates of its source node. To derive the repetition sequences, $\boldsymbol{v}$ is used to generate all the possible sequences that have to be XORed with the output of the source node to generate the output bit estimates of the SR node. Let $\eta_k$ denote the rightmost bit value of the left child node of the parent node of $\mathcal{N}^E_r$ at level $k+1$. When $v[k+1] = 0$, the left child node is a Rate-0 node so $\eta_k = 0$. When $v[k+1] = 1$, the left child node is a REP node, thus $\eta_k$ can take the value of either $0$ or $1$. The number of repetition sequences is dependent on the number of different values that $\eta_k$ can take. Let $W_{\boldsymbol{v}}$ denote the number of $1$'s in $\boldsymbol{v}$. The number of all possible repetition sequences is thus $2^{W_{\boldsymbol{v}}}$. Let $\mathbb{S} = \{\boldsymbol{s}_1,\ldots,\boldsymbol{s}_{2^{W_{\boldsymbol{v}}}}\}$ denote the set of all possible repetition sequences.

The output bits of SR node $\beta_i^j[1:2^j]$ have the property that their repetition sequence is repeated in blocks of length $2^{j-r}$. Let $\beta_r^E[1:2^r]$ denote the output bits of the source node of an SR node $\mathcal{N}_j^i$. The output bits for each block of length $2^{j-r}$ in $\mathcal{N}_j^i$ with respect to $\beta_r^E[1:2^r]$ can be written as
\begin{equation}
\label{eq:betaS}
\beta_j^i\left[\left(k-1\right)2^{j-r}+1:k2^{j-r}\right]=\beta_r^E\left[k\right]\oplus{\boldsymbol{s}_l},
\end{equation}
where $k\in\left\{1,\dots,2^r\right\}$ and $\boldsymbol{s}_l = \{s_l[1],\ldots,s_l[2^{j-r}]\}$ is the $l$-th repetition sequence in $\mathbb{S}$. To obtain the repetition sequence $\boldsymbol{s}_l$ and with a slight abuse of terminology and notation for convenience, the Kronecker sum operator $\boxplus$ is used, which is equivalent to the Kronecker product operator, except that addition in GF($2$) is used instead of multiplication. For each set of values that $\eta_k$'s can take, $\boldsymbol{s}_l$ can be calculated as
\begin{equation}
\boldsymbol{s}_l=\left(\eta_r,0\right)\boxplus\left(\eta_{r+1},0\right)\boxplus\cdots\boxplus \left(\eta_{j-1},0\right).
\label{eq:S}
\end{equation}
For a given code, the locations of SR nodes in the decoding tree are fixed and can be determined offline. Therefore, the repetition sequences in ${\mathbb{S}}$ of all of the SR nodes can be pre-computed and used in the course of decoding.

\subsection{Decoding of SR Nodes}

To decode SR nodes, the LLR values $\alpha_{r_l}^E[1:2^r]$ of the source node $\mathcal{N}_r^E$ associated with the $l$-th repetition sequence $\boldsymbol{s}_l$ are calculated based on the LLR values $\alpha_{j}^i[1:2^j]$ of the SR node $\mathcal{N}_j^i$ and repetition sequence $\boldsymbol{s}_l$ by the following equation which is proved in~\cite[Proposition 1]{sr2020}
    \begin{equation}
\label{eq:alphaS}    
\alpha_{r_l}^E\left[k\right]= \sum_{m = 1}^{2^{j-r}} \alpha_{j}^i\left[\left(k-1\right)2^{j-r}+m\right] \left(-1\right)^{s_l[m]}.
    \end{equation}

Using \eqref{eq:betaS} and \eqref{eq:alphaS}, \eqref{eq:estimate} can be written as~\cite[(20)]{sr2020}
\begin{equation}
\label{eq:estimate1}
\hat\beta_j^i=\mkern-10mu\underset{\substack{\beta_r^E\left[1:2^r\right]\in\mathbb{C}_r^E\\l\in \{1,\ldots,\left|{\mathbb S}\right|\}}}{\argmax}\sum_{k=1}^{2^r}\left(-1\right)^{\beta_r^E\left[k\right]}\alpha_{r_l}^E\left[k\right].
\end{equation}
Thus, the bit estimates of an SR node $\hat\beta_j^i\left[1:2^j\right]$ can be calculated by finding the bit estimates of its source node $\beta_r^E\left[1:2^r\right]$ and the repetition sequence using \eqref{eq:estimate1}, and then combine them as shown in \eqref{eq:betaS}.
\begin{algorithm}[t]
\caption{Decoding algorithm of SR node $\mathcal{N}_j^i$} 
\begin{algorithmic}
\REQUIRE $\alpha_j^i\left[1:2^j\right]$, ${\mathbb S}$;
\ENSURE $\hat\beta_j^i\left[1:2^j\right]$;\\
\tcp{Step 1: Soft message computation}
\For{$l\in\left\{1,\dots,\left|{\mathbb S}\right|\right\}$}
{\STATE Calculate $\;\alpha_{r_l}^E$ according to (\ref{eq:alphaS}).
}

\tcp{Step 2: Decoding of source node $\mathcal{N}_r^E$}
\For{$l\in\left\{1,\dots,\left|{\mathbb S}\right|\right\}$}{

\eIf{\text{SNT=Rate-C}}{Decode source node $\mathcal{N}_r^E$ using $\;\alpha_{r_l}^E$ to get $\hat\beta_{r_l}^E$.}{$\hat\beta_{r_l}^E\left[k\right]=h\left(\alpha_{r_l}^E\left[k\right]\right),\;k\in\left\{1,\dots,2^r\right\}$.\\
\If{$\text{SNT}\neq\text{Rate-1}$}{Perform parity check and bit flipping on $\hat\beta_{r_l}^E$ using $\alpha_{r_l}^E$.}
}
}

\tcp{Step 3: Comparison and selection}
\vspace{-0.25cm}
\begin{align}
\label{eq:comparison}
\hat l = \underset{l\in\left\{1,\dots,\left|\mathbb S\right|\right\}}{\argmax}\sum_{k=1}^{2^r}\left|\alpha_{r_l}^E\left[k\right]\right|.
\end{align}
Return $\hat \beta_{j_{\hat l}}^i$ to parent node according to (\ref{eq:betaS}).

\end{algorithmic}
\label{alg:alg1}
\end{algorithm}
%

The decoding algorithm of an SR node $\mathcal{N}_j^i$ is described in Algorithm~\ref{alg:alg1}. The algorithm first calculates $\alpha_{r_l}^E$ to obtain the soft messages that go into the source node for the $l^{th}$ repetition sequence ${\boldsymbol s}_l$, $l\in\left\{1,\dots,\left|{\mathbb S}\right|\right\}$. $\alpha_{r_l}^E$, $\widehat\beta_{r_l}^E$, and $\widehat \beta_{j_l}^i$ are the soft and hard messages associated with ${\boldsymbol s}_l$. Then, the source node is decoded under the rule of the SC decoding. If the source node is a special node, a hard decision is made directly. Parity check and bit flipping will be performed further using Wagner decoding if $\text{SNT}\neq\text{Rate-1}$. Finally, the index of the optimal repetition sequence can be selected according to the comparison in \eqref{eq:comparison} and the decoding result is obtained according to \eqref{eq:betaS}. Based on Algorithm~\ref{alg:alg1}, the SR node-based fast-SSC (SRFSC) decoding algorithm is proposed. It follows the SC decoding algorithm schedule until an SR node is encountered where Algorithm~\ref{alg:alg1} is executed.

\section{Architecture of SRFSC decoder}
\label{sec:Ar}
\begin{figure}[t]
\centering
\includegraphics[width=\columnwidth]{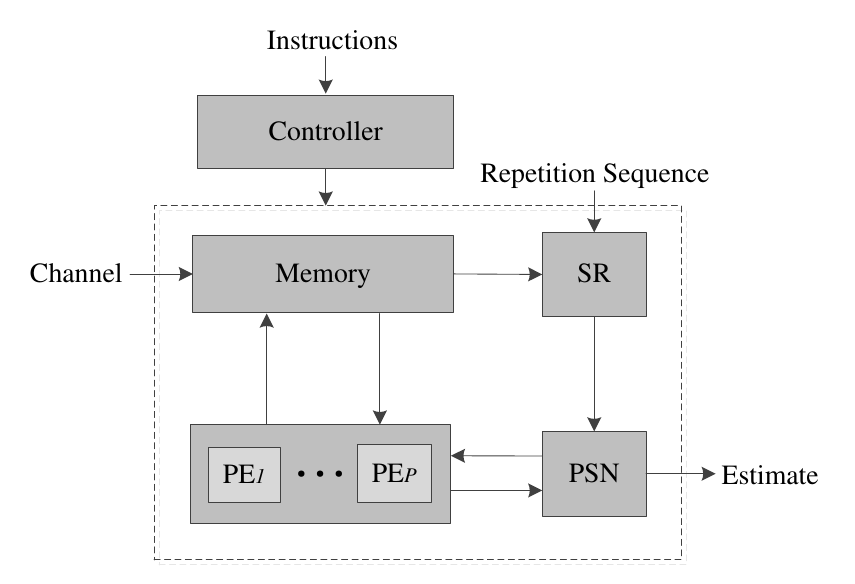}
\setlength{\abovecaptionskip}{-15pt}
\caption{Top-level architecture of the proposed SRFSC decoder.}
\label{fig:decoder}
\end{figure}

The top-level architecture of the proposed SRFSC decoder is shown in Fig.~\ref{fig:decoder}. When decoding starts, the  instructions for the polar code that is being decoded are fetched by the controller and the channel LLRs are loaded into memory. The controller decodes the instructions to get the node schedule and updates the decoding stage parameters accordingly. The updates in the controller follow the principle of SC decoding until an instruction corresponding to an SR node is reached, where the SR module is activated to process the LLRs. The estimation results from both the SR module and processing module are routed into the partial sum network (PSN) module, from where the estimated codeword is also output when decoding terminates. In the following, the architecture of the various individual modules is discussed in detail. 

\subsection{Memory, Processing, and PSN Modules}
The architectures of these three modules are identical with those presented in \cite{semi2013} and we thus only describe them on a high level. The memory module stores all soft messages $\alpha$. The update of hard estimates $\beta$ is in the partial sum network (PSN) module. A set of $P$ processing elements (PEs) is instantiated in the processing module to process up to $2P$ LLRs in parallel. A PE implements both the $f$ and the $g$ function using sign-and-magnitude representation and the appropriate output is selected according to the current decoding stage. 

\subsection{Controller Module}\label{sec:controller}
The operation in the controller module follows the standard SC decoding schedule until an instruction that indicates an SR node is found. When this occurs, the $2P$ LLRs will be routed to the SR module instead of the processing module to perform the decoding of SR node in Algorithm~\ref{alg:alg1}. The required number of clock cycles to decode the SR node by the SR module is pre-calculated and a counter is initialized to this value. All updates in the controller are suspended until the counter reaches zero. Then, the decoding bit index is added the length of the SR node and the updates resume. Although the Rate-0 and Rate-1 nodes can also be represented as special cases of SR nodes, the controller will bypass the SR module and signal the processing module to execute immediate decoding for these two nodes so that there is no additional latency.

\begin{figure}
\centering
\includegraphics[width=\columnwidth]{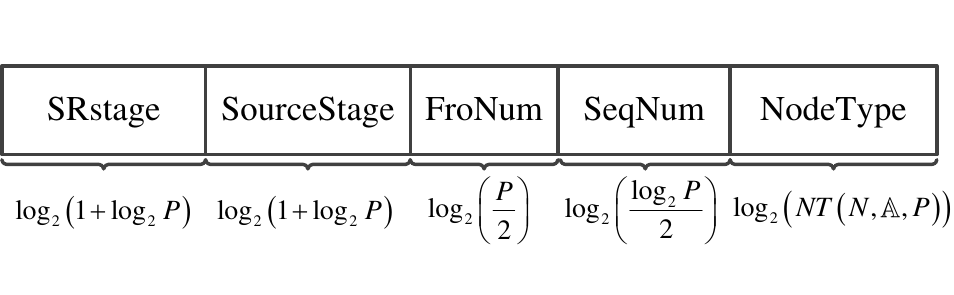}
\setlength{\abovecaptionskip}{-25pt}
\caption{Instruction structure of the proposed SRFSC decoder.}
\label{fig:schedule}
\end{figure}

The structure of the instructions used in the controller is shown in Fig.~\ref{fig:schedule}. The instructions contain all the required information to decode an SR node and they are stored in memory according to the visiting order in the decoding tree. The elements \texttt{SRstage}, \texttt{SourceStage}, \texttt{FroNum}, \texttt{SeqNum} and \texttt{NodeType} in the instruction represent the stage of SR node, the stage of source node, the number of frozen bits in source node, the base 2 logarithm of the number of repetition sequences and the node type, respectively.\footnote{Note that we use \texttt{FroNum} instead of SNT because no SR node with a Rate-C node as its source node is found for the code length $\left(N=1024\right)$ and rates $\left(R=1/2,\;1/4,\;3/4\right)$ that we consider in Section~\ref{sec:performance}}
Moreover, the vector $\boldsymbol{v}$ is replaced with \texttt{SRstage}, \texttt{SeqNum} and \texttt{NodeType} since these three elements can be used directly in the decoder, so that additional calculations (e.g., \eqref{eq:S}) can be avoided. \texttt{NodeType} is in fact a pointer to the memory of repetition sequences. As only nodes with $\texttt{SeqNum}>0$ have non-zero repetition sequences that need to be stored, \texttt{NodeType} refers to these node types and is used as pointer to find their corresponding repetition sequences in the memory.

The different repetition sequences in the SR node are processed in parallel. Since a maximum of $2P$ LLRs are input to the SR module each time, we have the constraint
\begin{equation}
2^{\texttt{SRstage}+\texttt{SeqNum}}\leq2P.
\label{eq:constraint}
\end{equation}
All SR nodes that meet this constraint can be handled, while others are divided into smaller nodes. Therefore, \texttt{SRstage} and \texttt{SourceStage} always have values between $0$ and ${1+\log_2 P}$. \texttt{FroNum} can be calculated according to \eqref{eq:a} and thus have values between 0 and $P/2$. Consider source node with a minimum length of 2. Then, the maximum value of \texttt{SeqNum} is constrained by $2^{1+2\texttt{SeqNum}}\leq2P$. 
Thus, \texttt{SeqNum} has values between $0$ and $\frac12\log_2 P$. As for \texttt{NodeType}, it has values between $0$ and $NT\left(N,\;\mathbb{A},\;P\right)$, where $NT$ is a function of $N$, $\mathbb{A}$ and $P$, which depends on the polar code being decoded.

As an example, we consider a set of 5G polar codes \cite{3GPP1} of length $N=1024$ and rates $R=1/2$, $R=1/4$, and $R=3/4$. For a code length of $N=1024$, $P=64$ is shown to be a reasonable choice \cite{Furkan2017}. With these parameters, in Fig.~\ref{fig:schedule}, \texttt{SRstage} and \texttt{SourceStage} take values in $\{0,1,\hdots,7\}$, \texttt{FroNum} takes values in $\{0,1,\hdots,3\}$, and \texttt{SeqNum} takes values in $\{0,1,2\}$. The three considered codes contain a total of six SR nodes with $\texttt{SeqNum}>0$. As such, \texttt{NodeType} takes values in $\{0,1,\hdots,6\}$. Specifically, when $\texttt{NodeType}=0$, the node only has an all-zero repetition sequence and the remaining values represent the six SR nodes with $\texttt{SeqNum}>0$. From the above analysis, the size of each instruction for the considered example is $13$ bits.

\begin{figure*}
\centering
\scalebox{0.8}{\includegraphics{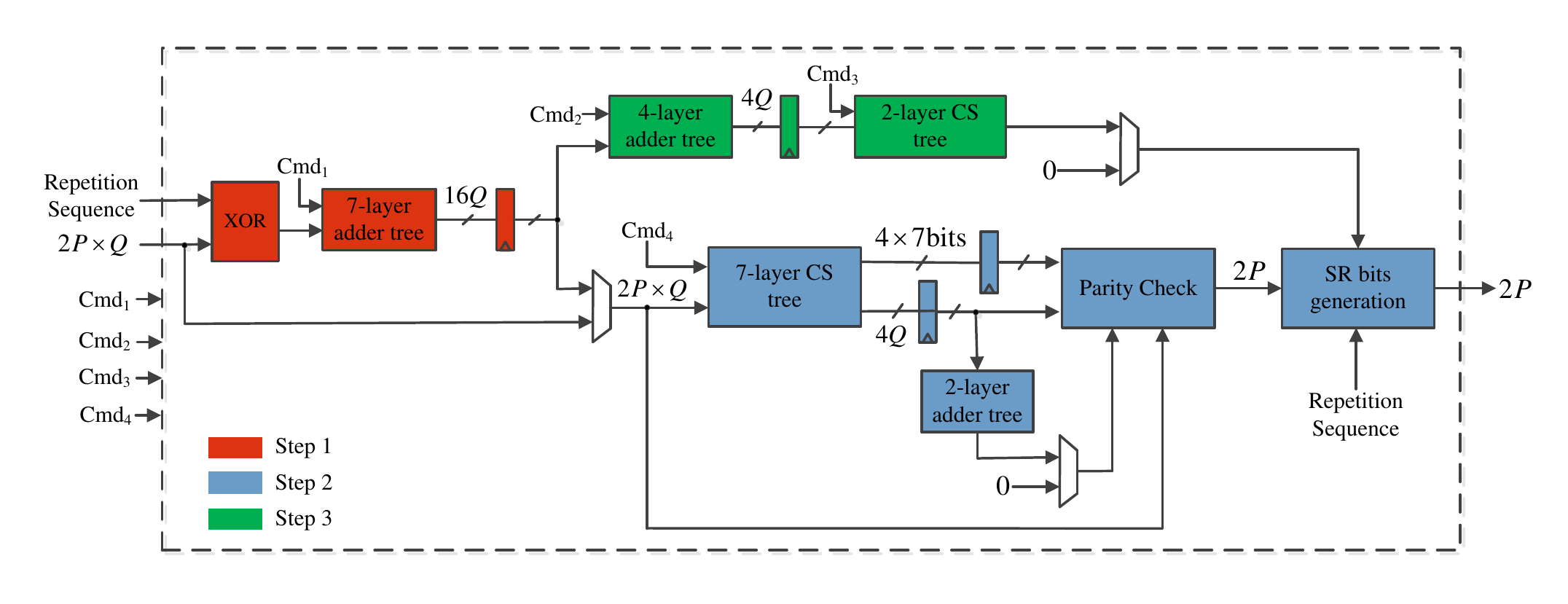}}
\setlength{\abovecaptionskip}{-15pt}
\caption{Example of the SR module architecture for $N=1024$, $R \in \{1/2,1/4,3/4\}$, and $P=64$.}
\label{fig:SRmodule}
\end{figure*}

\subsection{SR Module}

The ranges of some elements in the instructions are variable and depend on the set of supported polar codes. Thus, some of the data widths in the SR module are also variable and it is difficult to give a fully generic explanation of our proposed architecture. For this reason, we consider the previous example of $N=1024$, $R \in \{1/2,1/4,3/4\}$, and $P=64$. The architecture of the SR module for this example is shown in Fig.~\ref{fig:SRmodule}. The submodules with red, blue, and green color correspond to the operations in Step 1, Step 2, and Step 3 in Algorithm~\ref{alg:alg1}, respectively, and are explained in more detail in the sequel.
 
\textbf{\emph{Step 1:}} This part of the SR decoder is used to calculate the input LLRs into the source node if $\texttt{SRstage}\neq \texttt{SourceStage}$. In the XOR submodule, the first $2^{\texttt{SRstage}}$ LLRs in the $2P$ inputs are repeated $2^{\texttt{SeqNum}}$ times so that the decoding for different repetition sequences can be handled in parallel. The repetition sequences are obtained using \texttt{NodeType}. They will be XORed with the sign bit of the $2^{\texttt{SeqNum}}$ input repetitions according to \eqref{eq:alphaS}. The XOR result of different repetition sequences are concatenated and expanded into a vector of length $2P$ by appending  zeros if $2^{\texttt{SRstage}+\texttt{SeqNum}}<2P$. 
Then, the LLR vector enters a $\left(1+\log_2 P\right)$-layer adder tree that performs the addition of LLRs in~\eqref{eq:alphaS}. The command signal $\text{Cmd}_1=7-\left(\texttt{SRstage}-\texttt{SourceStage}\right)$ is pre-calculated in the control module and it is used in the adder tree to decide the addition result of which layer will be output by a multiplexer. Those outputs from the adder tree are the input LLRs of the source node for different repetition sequences. In the considered example, there exist $2^{\texttt{SourceStage}+\texttt{SeqNum}}\leq16$ for SR node whose $\texttt{SRstage}\neq \texttt{SourceStage}$. Moreover, all LLRs are quantized using $Q$ bits. Thus, the data width of the adder tree output is $16Q$ bits.

\textbf{\emph{Step 2:}} This part of the SR node is used to perform the parity-check and bit-flipping steps for the source node. The LLRs of the source node first enter a {$\left(1+\log_2 P\right)$-layer} compare-select (CS) tree. Processing units in the CS tree execute the $f$ function to decode SPC node. There are two cases where more than one SPC nodes will be decoded in parallel in our design: 1) when $\texttt{FroNum}=1$ and $\texttt{SeqNum}>0$, there are $2^{\texttt{SeqNum}}$ SPC nodes which correspond to different repetition sequences and are decoded simultaneously, and 2) when $\texttt{FroNum}=2$ and $\texttt{FroNum}=3$, the decoding of source node can be viewed as a parallel decoding of 2 and 4 SPC nodes, respectively \cite{sr2020}. The length of the SPC node decides the layer from which the index of the least reliable input and the $f$ function result are selected. As the length of the SPC node can be calculated as $2^{\texttt{SourceStage}+1-\texttt{FroNum}}$, the output layer selection signal Cmd$_4$ has the following representation 
\begin{align}
\text{Cmd}_4=\left\{ 
    \begin{matrix}
        7, &\texttt{FroNum}=0,\\
        6-\texttt{SourceStage}+\texttt{FroNum}, & \text{otherwise}.
    \end{matrix} 
    \right.
\end{align}
Since the maximum number of parallel SPC nodes in our example is 4, the output indices and LLRs have a data width of $4\times7$ and $4Q$ bits, respectively. Note that the output LLRs goes both to the parity check module and a 2-layer adder tree. This is because all SPC nodes have an even parity constraint except when $\texttt{FroNum}=3$, where SPC nodes can have an even or odd parity constraint which is calculated according to~\cite[(16)]{sr2020} and implemented by a 2-layer adder tree.  

The parity constraint type, the output indices, and LLRs are then input into the parity-check submodule to do the parity check and bit flipping on these SPC nodes using~\cite[(13)]{sr2020}. Then, the estimated bits of these SPC nodes are concatenated to form the estimated bits of source node and they are XORed with the repetition sequence to generate the estimated bits of SR node in the SR bits generation submodule according to \eqref{eq:betaS}. Finally, the SR bits corresponding to the repetition sequence with the index value from Step 3 are selected as the output.

\textbf{\emph{Step 3:}} This part of the SR decoder is executed in parallel with Step 2 to evaluate~\eqref{eq:comparison} using a {$\left(\texttt{SourceStage}+\texttt{SeqNum}\right)_{\max}$-layer} adder tree and $\texttt{SeqNum}_{\max}$-layer CS tree, where $\left(\texttt{SourceStage}+\texttt{SeqNum}\right)_{\max}$ is the maximum value of $\left(\texttt{SourceStage}+\texttt{SeqNum}\right)$ for all SR nodes with $\texttt{SeqNum}>0$ and $\texttt{SeqNum}_{\max}$ denotes the maximum value of $\texttt{SeqNum}$. 
As only magnitudes of LLRs are used for addition in \eqref{eq:comparison}, all inputs are positive. As a result, the processing unit in the 4-layer adder tree is simpler than that in the 7-layer adder tree in Step 1 because it does not need to compare magnitudes. The output of the adder tree is selected by the output layer selection signal Cmd$_2=4-\texttt{SourceStage}$ and has a bit-width of $4Q$ as there are at most 4 repetition sequences in the considered example. The four sums are then input into the 2-layer CS tree to find the index of the maximum using selection signal Cmd$_3=2-\texttt{SeqNum}$. Finally, the index is obtained from a multiplexer and the value is 0 if $\texttt{SeqNum}=0$ and the output from the CS tree otherwise.

\section{Implementation Results}
\label{sec:performance}

\begin{figure}
  \leftline{ 
\includegraphics{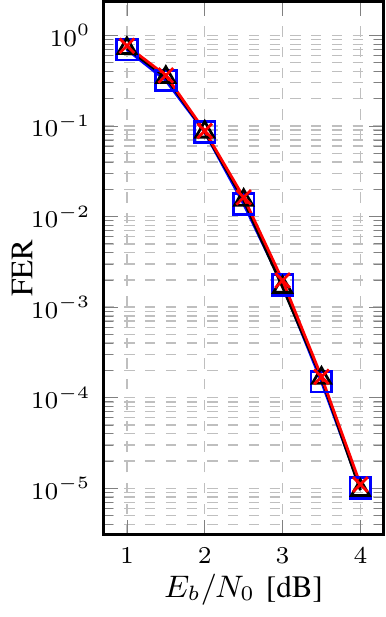}
\includegraphics{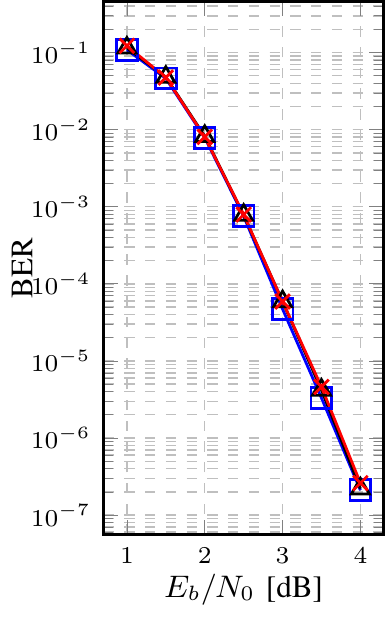}}
\centering
\includegraphics{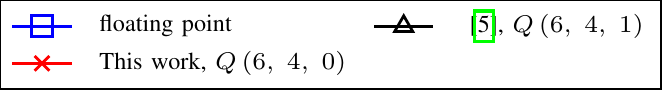}
  \caption{Floating-point and fixed-point FER and BER performance for SRFSC decoding of 5G polar codes $\mathcal{P}\left(1024,512\right)$ \cite{3GPP1}.}
  \label{fig:SR-performance}
\end{figure}

The proposed decoder has been implemented using VHDL and targeting an Altera Stratix IV EP4SGX530KH40C2 FPGA device. Channel LLRs are generated by transmitting random codewords through an additive white Gaussian noise (AWGN) channel after binary phase-shift keying (BPSK) modulation. A quantization scheme $Q\left(6,\;4,\;0\right)$ has
been used, where $Q\left(Q_i,\;Q_c,\;Q_f\right)$ are the quantization bit size
for internal LLRs, channel LLRs, and fraction bit size for
both internal and channel LLRs, respectively. This scheme leads to an error-correcting performance that is very close to that of the floating-point implementation, as shown in Fig.~\ref{fig:SR-performance}. 

Table~\ref{tab:com2} compares the proposed decoder with other state-of-the-art works. As can be seen, the proposed SRFSC decoder provides a $17.9\%$ and $31.7\%$ throughput improvement compared to the architectures presented in \cite{Furkan2017} and \cite{sarkis2014fast}, respectively. This is mainly due to a $9.9\%$ and $22.4\%$ higher $f_{max}$ with respect to \cite{sarkis2014fast} and \cite{Furkan2017}. The number of CLKs in our work is slightly higher than that in \cite{Furkan2017}. This is because of the insertion of some registers to decrease certain critical paths and because we have not merged $f$ ad $g$ operations as was done in~\cite{Furkan2017}. In addition, a total of 186, 200 CLKs are required at rates $1/4$, $3/4$, respectively. In terms of the used LUTs, this work requires an increase of $23.2\%$ and $187.5\%$ compared to \cite{Furkan2017} and \cite{sarkis2014fast}, respectively. As far as the memory size is concerned, although our decoder uses fewer RAM bits, the required number of registers is about 8 times higher compared to~\cite{Furkan2017,sarkis2014fast}. The big difference in registers can be mostly attributed to the separate storage of channel and internal LLRs in synthesis. Internal LLRs are stored in RAM and channel LLRs are arranged in registers, while in other works both are stored in RAM.

\begin{table}[t] 
\centering
\caption{FPGA Implementation Results for $\mathcal{P}\left(1024,512\right)$.}
\renewcommand\arraystretch{1.25}
        \begin{tabular}{p{2.0cm}<{\centering}p{1.5cm}<{\centering}p{1.5cm}<{\centering}p{1.5cm}<{\centering}}
            \toprule
             & \cite{sarkis2014fast} & \cite{Furkan2017} & This Work \\
            \midrule
            Quantization & $Q\left(6,\;4,\;1\right)$ & $Q\left(6,\;4,\;1\right)$ & $Q\left(6,\;4,\;0\right)$\\
            $P$ & 64 & 64 & 64 \\
            LUTs & \textbf{6126} & 14300 & 17615 \\
            Registers & 1223 & \textbf{1216} & 10505 \\
            RAM (bits) & 23592 & 18350 & \textbf{16128} \\ 
            Instruction size & \textbf{5} bits & 6 bits & 13 bits \\ 
            $\#$ of Instruction & 209 & 157 & \textbf{41} \\          
            $\#$ of CLKs & 266 & \textbf{214} & 222 \\ 
            $f_{\max}$ (MHz) & 99.8 & 89.6 & \textbf{109.6} \\
            $T/P$ (Mps) & 384 & 428.6 & \textbf{505.6} \\             
\bottomrule
        \end{tabular}
\label{tab:com2}   
\end{table}

\section{Conclusion}
\label{sec:Conclu}

In this paper, we presented the first FPGA implementation of the SRFSC decoder for polar codes. 
To this end, we designed a dedicated architecture for the SR node processor. For a 5G polar code 
with length 1024, code rate $1/2$ and $P=64$ processing units, we obtained a $17.9\%$ improvement in throughput over the previous work.

\ifCLASSOPTIONcaptionsoff
  \newpage
\fi

\bibliography{IEEEabrv,myreference}

\end{document}